\RequirePackage{lineno}
\documentclass[prb,twocolumn,aps,amsmath,amssymb,floatfix]{revtex4}
\usepackage{graphicx} 
\usepackage{bm} 
\usepackage{ulem}
\usepackage{color,graphicx,pstricks}

\begin{document}

\title{Itinerant ferromagnetism in a spin-fermion model for diluted spin systems}

\author{Sourav Chakraborty$^{}$, Sandip Halder$^{}$ and 
Kalpataru Pradhan$^{}$\footnote{kalpataru.pradhan@saha.ac.in}}
\affiliation{$^{}$Theory Division, Saha Institute of Nuclear Physics, 
HBNI, Kolkata-700064, India}

\date{\today}

\begin{abstract}
We investigate the itinerant ferromagnetism using a diluted spin-fermion model,
derived from a repulsive Hubbard model, where itinerant fermions are coupled
antiferromagnetically to auxiliary fields in a three-dimensional simple cubic
lattice. We focus, in particular, on understanding the spin-dependent transport
properties of the itinerant fermions in the impurity band by taking positional
disorder of the auxiliary fields into account. For on-site repulsion $U$ $\sim$
bandwidth the density of the itinerant carriers confined to the impurity band,
play a key role in determining the kinetic energy of the system and
consequently the carrier spin polarization. Our
semi-classical Monte Carlo calculations show that the ferromagnetic transition
temperature of the carrier spins indeed shows an optimization behavior with the 
carrier density. We calculate the transport properties in details to establish 
a one-to-one correspondence between the magnetic and transport properties of
the carriers. Our results obtained beyond the perturbative regime are significant
for understanding the ferromagnetism in diluted magnetic semiconductors.
\end{abstract}

\maketitle 

\section{INTRODUCTION}

The history of magnetism is very very old but scientists have begun to understand
the concept during the twentieth century~\cite{buschow}. A detailed understanding
of magnetism is always necessary to invent and design new magnetic
materials~\cite{fert,gruenberg,ohno,spladin}. As of now microscopic description of itinerant
ferromagnetism still remains a subject of intense research~\cite{prange,sakai,massignan,holzmann,jo}.
Due to the technological applications even today the study of itinerant ferromagnetism
remains as one of the most interesting
as well as challenging avenue for both experimental and theoretical condensed
matter physicists.

In recent years, significant progress has been made to understand the magnetic ordering
using the microscopic theories of itinerant magnetism.
In 1938, Stoner model was first introduced to interpret the itinerant ferromagnetism~\cite{stoner}.
In his phenomenal work Stoner pointed out that the ferromagnetic order can arise 
due to the interaction among the itinerant electrons, which spin spilt the 
electronic band structure. For various metals, such as Fe, Co, and Ni, the
itinerant electrons exhibit ferromagnetic behavior~\cite{stoner,korenman}. 
In these materials electrons whose spins aligned to form ferromagnetic state 
are extended and give rise to metallicity. Earlier to Stoner, 
one of the first justifications of quantum ferromagnetism was put forward by 
Heisenberg who established that the exchange interactions, arising due to 
spin-dependent Coulomb repulsion, drives the magnetism between localized
moments~\cite{heisenberg}.

Heisenberg model both in classical and quantum forms remains one of the finest
and oldest tools to explain the magnetism and related physical observables in 
strongly correlated magnetic insulators~\cite{matthias,mauger,anderson,anderson1}.
Although ferromagnetic Heisenberg model is explored to understand itinerant
ferromagnetism~\cite{mattis} but the real difficulty lies in arriving at the
effective Hamiltonian of an interacting spin system with negative exchange coupling~\cite{anderson}.
Ferromagnetic kinetic exchange between localized spins that arises from an interplay
of spin and orbital degrees of freedom is relatively rare~\cite{huang}.

However, probing ferromagnetism in diluted ferromagnetic semiconductors using 
Heisenberg model is limited~\cite{bergqvist,hilbert,barzykin}. Dual semiconducting and 
magnetic property of ferromagnetic semiconductors~\cite{munekata,ohno1}, an diluted spin system, 
where magnetic impurities are doped in a host semiconductor, is expected to bring 
technological revolution in spintronics~\cite{ohno2,dietl,dietl1}.
Theoretical investigation is necessary to understand the physics of these materials
that will help the us to push the ferromagnetic transition $T_C$ beyond the room
temperature~\cite{jungwirth}.

In this class of diluted spin systems the magnetic impurities provides both the
itinerant carriers and localized moments~\cite{dietl1,jungwirth}. The itinerant carriers
reside in the shallow acceptor level introduced by magnetic impurity ions in the
host semiconductor band gap. It is widely accepted that the (exchange) interaction
between the magnetic
spins is mediated by the itinerant electrons~\cite{zener1,zener2, dietl2,macdonald,calderon}. This warrants an additional
inter-band coupling between the itinerant carriers and the localized moments to
study the magnetic and transport properties of charge carriers.

In order to understand
the physics of spin-spliting in the carrier impurity band from the perspective 
of {\it itinerant-exchange} mechanism we focus on strong coupling limit.
In this limit carriers are firmly localized to the impurity
sites and as a result the acceptor levels give rise to distinct impurity band~\cite{sanvito,mahadevan}.
Although impurity band picture in the most studied GaMnAs semiconductor remains controversial
till date~\cite{samarth,hirakawa,sapega,dobrowolska}, addition of Mn to GaN like large band gap semiconductors give
rise to a deep impurity band within the host band gap~\cite{kronik,korotkov,bouzerar}. Due to disorder
(anti-site disorder~\cite{myers} and interstitial defects~\cite{yu}) the impurity band
remains less than half filled, i.e. carrier density remains smaller than the impurity
density and the Fermi energy lies in the impurity band. The position of the Fermi energy,
decided by the density of the itinerant carriers plays a key role in determining
the kinetic energy of the system. Consequently, the gain in kinetic energy is considered
a major factor that decides the carrier spin polarization. Detailed study comprising of
carrier-spin-dependent transport of localized carriers are limited to date.

In this work we focus on the spin-dependent transport properties of the carriers 
confined to the impurity band. In our effective spin-fermion model Hamiltonian, 
derived from repulsive Hubbard model, we assign the $U$ $\sim BW$ (band width) on 
few percentage of sites in a simple cubic lattice and set it to zero for rest of 
the sites. We take the carrier density with respect to the impurity concentration 
which is concomitant with experimental measurements. We organize this paper
as follows: In section II we introduce the effective spin-fermion model derived
from the Hubbard Hamiltonian and outline our method. We frame the impurity
band scenario in Section III. In Section IV we present our
numerical results comprising of spin-dependent transport of carriers for 
$U$ $\sim$ BW. Section V is dedicated to compare the magnetic and transport
properties by varying the on-site interaction $U$.
Section VI is devoted to analyze our main results for two different
concentration of the impurities. Finally in section VII we summarize our results.

\section{MODEL HAMILTONIAN AND METHOD}

We consider one band electron-hole symmetric Hubbard Hamiltonian
{\small
\begin{eqnarray}
H= -t\sum_{<i,j>,\sigma} (c^{\dagger}_{i,\sigma}c_{j,\sigma}+h.c.)
 + U\sum_{i} \Big(n_{i,\uparrow}-\frac{1}{2}\Big)
  \Big(n_{i,\downarrow}-\frac{1}{2}\Big) \nonumber
\end{eqnarray}}

\noindent
where first term is the kinetic energy [$t$ is the nearest neighbor hopping
parameter and $c^{\rm \dagger}_{i\sigma}$ ($c_{i\sigma}$) are the fermion
creation (annihilation) operators at site $i$ with spin $\sigma$] and the
second term presents the repulsive Hubbard interaction ($U$ $>$ 0). 

We reduce the quartic fermion problem present in repulsive Hubbard model into 
a quadratic
one by introducing Hubbard-Stratonovich field and extract the following effective
spin-fermion type Hamiltonian by suppressing the imaginary-time dependence from
the Hubbard-Stratonovich fields
(for details please see Refs.~[\citenum{mukherjee,chakraborty1}]). Then above
Hamiltonian transfers to

{\small
\begin{eqnarray}
H_{sf}=-t\sum_{<i,j>,\sigma} (c^{\dagger}_{i\sigma}c_{j\sigma} + h.c.)\nonumber
+ U/2 \sum_{i} (\textless n_i \textgreater n_i - \textbf{m}_i . \sigma_{i}) \nonumber \\ 
+ (U/4)\sum_{i}(\textbf{m}_{i}^2 - {\textless n_i \textgreater}^2)
-\frac{U}{2} \sum_{i} n_{i} \nonumber
\end{eqnarray}}

\noindent
where fermions are coupled to the classical auxiliary fields ($\bf m_i$).

The non-monotonic $U$ dependence of the antiferromagnetic transition temperature $T_N$ is
established for undiluted system ($U$ at all sites, i.e. $x=1$ limit) at half filling
using a semi-classical Monte Carlo (s-MC) approach~\cite{mukherjee} that matches well 
with DQMC results~\cite{ulmke}. Recently, we have used s-MC approach to show that
the antiferromagnetic order persists beyond the classical percolation threshold in
the diluted one-band Hubbard model at absolute half filling in three
dimensions~\cite{chakraborty1}.

For diluted spin systems, we have considered finite $U \sim BW$ at randomly chosen
sites {\it k} (with concentration $x$) and put $U=0$ for rest of the sites (with concentration
$1-x$). Our diluted Hamiltonian is of the form:
{\small
\begin{eqnarray}
H_{sf}= -t\sum_{<i,j>,\sigma} (c^{\dagger}_{i,\sigma}c_{j,\sigma}+ h.c.)
+ U/2 \sum_{k} (\textless n_k \textgreater n_k
 - \textbf{m}_k . \sigma_{k})  \nonumber \\
  + (U/4)\sum_{k}(\textbf{m}_{k}^2 - {\textless n_k \textgreater}^2)
   -\frac{U}{2} \sum_{k} n_{k}  -\mu \sum_{i} n_i \nonumber
\end{eqnarray}}

\noindent
The overall electron density $n$ is controlled through the chemical potential 
($\mu$) given in the last term. $\mu$ is chosen self-consistently during the 
thermalization process to get the desired electron density $n$ at each
temperature. In quantum Monte Carlo method $\textbf{m}_{k}$
variables are often used as Ising-like~\cite{qin,shi,bouadim}.
These Ising-like auxiliary fields were introduced by Hirsch through a discrete
Hubbard-Stratonovich transformation~\cite{hirsch}. So, we use Ising-like
auxiliary fields in our calculations. For this case the carriers
are either point towards up or down direction in strong coupling limit.

We use semi-classical Monte Carlo (s-MC) method to anneal
the system from high temperature consisting of randomly oriented auxiliary 
fields to obtain the ground state for a fixed carrier density. First we 
chose a set of random auxiliary fields {$\textbf{m}_{k}$} at desired number
of sites and set {$\langle n_k \rangle$} to be uniform for a system size of
N = $L^3$ = $10^3$ and calculate the internal energy of the carriers by
exact diagonalization scheme. Then we update the auxiliary field
at an impurity site say $k$ and recalculate the internal energy of the carrier
using the new auxiliary field configuration. We employ Metropolis algorithm
to accept or reject the above update. At every 10th Monte Carlo step, using
the resulting {$\textbf{m}_k$} configuration, we update {$\langle n_k \rangle$} 
self-consistently. This new set of {$\langle n_k \rangle$} is used to 
perform the further Monte Carlo steps. In order to access large system size
we adopt a Monte Carlo update technique based on the traveling cluster
approximation (TCA)~\cite{sanjeev,chakraborty}.

For impurity concentration $x$ we assign finite $U$ for $10^{3}x$ sites (for
a system size $10^3$) randomly and set $U = 0$ for rest of the sites. We define
carrier density as the electrons (or holes) per impurity site. Direct 
exchange interaction between the impurity spin sites is not taken in to
account, which is a valid approximation in the diluted limit, by avoiding the
nearest-neighbor impurity pairing. All physical quantities such as carrier
magnetization and conductivity are averaged over ten different positional
disorder configurations of auxiliary fields in addition to the quantum and
thermal averages taken during the Monte Carlo simulations. In this work we
consider mainly $x=0.25$, but compare our main results between $x=0.25$ and
$0.125$ at the end. All parameters such as on-site repulsion ($U$) and temperature
($T$) are scaled with hopping parameter ($t$). We use U$\sim$BW and vary the
carrier density with respect to $x$ from $0$ to $1$.

\begin{figure}[t]
\centering

\includegraphics[scale=0.35]{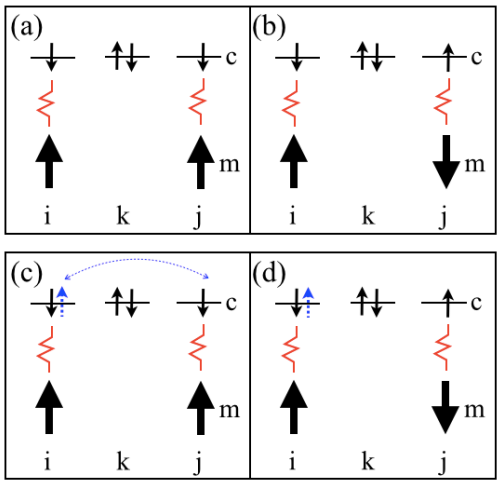}
\caption{Schematic: Large arrows indicate the auxiliary fields while small 
arrows are the carriers (shown for sites $i$ and $j$). We also shows the 
intervening non-impurity site ($k$). 
Top Panel: Half filling (one carrier per each impurity site) case. We find 
that the carriers orient randomly due to intervening non-impurity site at 
half filling. None out of (a) and (b) are found to be the ground state.  
Bottom Panel: Our calculations shows ferromagnetic state is favored due to 
gain in kinetic energy via the non-impurity site as shown in (c) when compared to 
the scenario drawn in (d).}
	
\label{fig01}
\end{figure}

\section{Impurity Band Picture}

In the impurity band picture, which is relevant to ferromagnetic semiconductors,
carriers reside in the shallow acceptor level separated from the valence band due to the
strong coupling between the impurity ions and the carriers. The width of the impurity
band and gap between impurity band and valence band depend upon the coupling strength. 
The location of the Fermi energy inside the impurity band plays a vital role in determining
the transport and magnetic properties of the system. The optimum $T_C$ is expected 
for which Fermi energy is at the center of the band to gain maximum kinetic energy 
from the delocalization of the carriers and supposed to decrease towards the edge 
of the impurity band.

\begin{figure}[t]
\centering

\includegraphics[scale=0.3]{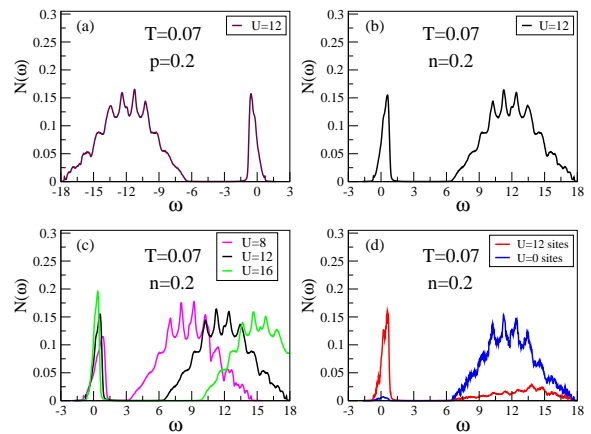}
\caption{Density of states for carrier density $p$ or $n$ = 0.2 at fixed $x=0.25$
depicts a distinct impurity band for $U \sim BW$. DOS for (a) hole density 
$p$ = 0.2 and (b) electron density $n$ = 0.2 for $U = 12$. (c) The gap between
the impurity band and the valence band increases with increasing the on-site
interaction $U$ and the IB shrinks. (d) Site resolved DOS for $U = 12$ case. 
All the density of states are plotted for $T=0.07$. The Fermi energy is set at zero.}

\label{fig02}
\end{figure}

Using our semi-classical Monte Carlo calculations we identify that the itinerant
ferromagnetic order obtained in the impurity band picture in spin-fermion model is
due to the following scenario of event in which one type of carrier (say up)
is more mobile than the other one (down) and drives the magnetism. A schematic
figure is shown in Fig.~\ref{fig01} by assuming an
impurity band picture where carriers are antiferromagnetically aligned to auxiliary
fields at the impurity sites (sites $i$ and $j$). Intervening non-impurity site $k$ 
is also shown for completeness. For half-filling case (one carrier per each impurity site)
one would naively expect an AFM ground state mediated by carrier via the non-impurity
sites. But it is also apparent the magnetic
ground state is strongly depends on the magnetic impurity concentration. For $x=0.25$,
which is below the classical percolation limit ($x_p^{sc}$$\sim$0.31), we do not find any magnetic
ordering. This shows that the antiferromagnetic coupling between the auxiliary fields
and localized carriers favors the paramagnetism at half filling. Beyond half filling
case extra electron added to the system is now relatively more mobile as the first
electron is already anti-aligned with auxiliary field and give
rise to the ferromagnetic ground state by maximizing the kinetic energy via the 
non-impurity site(s).

We start our calculation for $x=0.25$ and use $U = 12$ ($\sim BW$) to
manifest the formation of an impurity band that imposes the carrier
localization. A well separated impurity band for hole density $p=0.2$ is clear
from density of states (DOS) shown in Fig.~\ref{fig02}(a) for a relatively high temperature ($T = 0.07$).
The DOS at each $\omega$ is obtained by implementing the Lorentzian representation
of the $\delta$ function:
$N(\omega) = \sum_{k} \delta(w - w_{k})$, where $\omega_{k}$ are the eigenvalues
of the fermionic sector and the summation runs up to total number
of eigenvalues ($2L^{3}$) of the system. The valence band is very much
symmetric but the narrow impurity band is asymmetric. This asymmetric character
of the impurity band picture remains intact for all the carrier densities. It is
important to mention here that ferromagnetism in Hubbard model is attributed to
an asymmetric density of states with large spectral weight near one of the
band edges~\cite{ulmke1,oberm,kollar}.

For high band gap systems the ferromagnetism along with the nature of the charge
carriers is controversial till date. Holes (electrons) are accounted to be the
charge carriers for p-type conduction~\cite{edmonds,dietl1} (n-type
conduction~\cite{coey,xu}) in ferromagnetic semiconductors. In Fig.~\ref{fig02}(b)
we also plot the DOS using the electron density ($n = 0.2$). The structure of the
impurity band and the position of the Fermi energy for both hole ($p$) and
electron ($n$) pictures indicate that the magnetic and transport properties
would provide very similar results in our s-MC calculations,
which is expected from a particle-hole symmetric model Hamiltonian. So,
for brevity we performed all our calculations by varying electron density $n$.

\begin{figure}[t]
\centering

\includegraphics[scale=0.33]{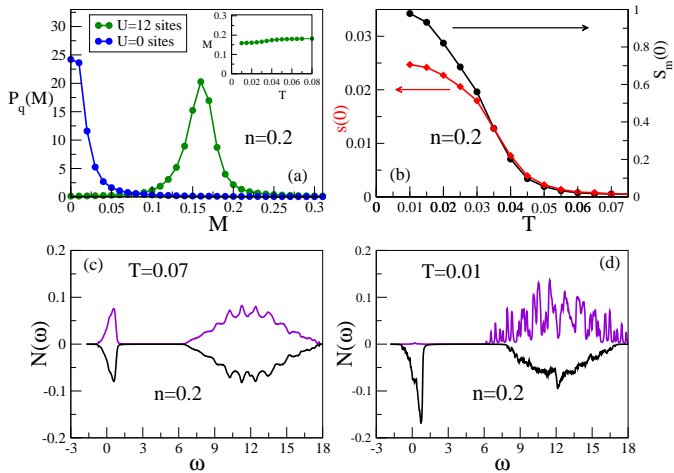}
\caption{Carrier spin polarization for $n$=0.2 (using $U$=12 and $x=0.25$):
(a) The distribution of carrier moments $P_q(M)$ for $U = 12$ (distribution
for $U = 12$ sites and $U = 0$ sites are plotted separately).
(b) Quantum $s({\bf 0})$ and classical $S_m({\bf 0})$ ferromagnetic structure factor 
vs temperature show the same transition. The spin resolved density of states 
are plotted for (c) $T = 0.07$ (above $T_C$) (d) $T = 0.01$ (below $T_C$). 
The equal contribution from up or down spin sector in (c) implies that the bands are 
unpolarized. For $T = 0.01$ the impurity band is completely spin polarized.
The Fermi energy is set at zero.}

\label{fig03}
\end{figure}

We plot the DOS for three different values of $U$ in Fig.~\ref{fig02}(c). The IB gets narrower
and the gap between the valence band and impurity band increases with increasing $U$.
This indicates that the carriers tend to be more and more localized in a finite
region of the lattice comprised of the impurities. In order to substantiate this
fact we plot DOS for $U=12$ sites and $U=0$ sites separately in Fig.~\ref{fig02}(d).
It shows that $U=12$ sites mostly contribute to the formation of the impurity band
whereas $U=0$ sites give rise to the valence band. The contribution of $U=12$ sites
in the valence band is due to the leaking of small amount of carriers from the
impurity sites to the host lattice. This contribution decreases upon increasing the
$U$ values.

\section{Spin dependent transport properties of carriers}

In order to analyze the carrier magnetism first we calculate the magnetic moments
$M$ [$M = \langle (n_{\uparrow} - n_{\downarrow})^{2} \rangle =                  
\langle n \rangle - 2 \langle n_{\uparrow} n_{\downarrow} \rangle$, where the
angular brackets imply quantum and thermal averaging] on each impurity ($U=12$) sites.
The system averaged quantum local moments at $U=12$ sites for $n=0.2$ is plotted in
inset of Fig.~\ref{fig03}(a). The average moments is approximately equal to the
carrier density barring the small change that is because of the carrier leakage to
the host band as discussed in Fig.~\ref{fig02}(d). Due to the carrier localization,
for $U \sim BW$, it is expected that the formation of moments at impurity site to
be more or less uniform.
The local moment distribution $P_q(M) = \sum_{M_i} \delta(M-M_i)$ at $T=0.01$ in
Fig.~\ref{fig03}(a) for $U = 12$ sites depicts this fact. The moments distribution
for $U=0$ sites, plotted separately in the same figure which shows that the moment
formations in the host lattice is minimal.

\begin{figure}[t]
\centering

\includegraphics[scale=0.3]{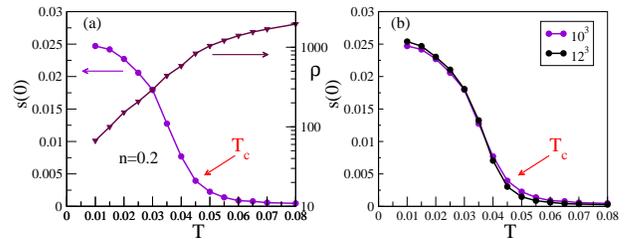}
\caption{Correlating ferromagnetism and metallicity for $n=0.2$
(for $U=12$ and $x=0.25$):
(a) Temperature dependence of resistivity in units of $\hbar a/\pi e^{2}$ 
(a is the lattice constant) shows metallicity at low temperature. Quantum 
structure factor $s(0)$ is also re-plotted in the same figure to draw a 
correspondence between the metallicity and the ferromagnetism.
(b) Quantum ferromagnetic structure factor $s(0)$ with temperature for 
two system sizes using $L=10$ and $12$.}

\label{fig04}
\end{figure}

Using carrier magnetic moments we now calculate the quantum ferromagnetic
structure factor $s(0)$ [where $s(\bold{q})=\frac{1}{(Nx)^2} \sum_{ij} \bold{M}_{i}.\bold{M}_{j}                       
  e^{i\bold{q}.(r_{i}-r_{j})}$]. This quantum observable is calculated by
using the eigenvectors resulted from exactly diagonalizing the equilibrated field
configurations. As the quantum structure factor involve four fermionic operators
Wick's theorem is used to transform the four fermionic expectation value to
combinations of two fermionic expectation values~\cite{dagotto}. 
We plot quantum ferromagnetic
structure factor $s(0)$ with temperature in Fig.~\ref{fig03}(b). For
$n=0.2$ maximum value of $s(0)$ can be 0.04 (if one gets perfect moment i.e.
$M$=0.2 at each $U=12$ sites). Although this is not the case here,
as shown in inset of Fig.~\ref{fig03}(a), there is a clear ferromagnetic transition.

We present the spin resolved DOS for both high and low temperature cases in
Fig.~\ref{fig03}(c) and (d). For $T=0.07$ (which is above $T_C$) the impurity band remain
unpolarized. Both the valence band and the impurity band are completely symmetric for
both up or down spin sectors. The impurity band is completely spin polarized for
$T=0.01$ which depicts the complete ferromagnetic ordering of the carriers that
reside within the impurity band. This agrees well with experiment~\cite{ohya}.
In addition we plot the classical structure
factor for the auxiliary fields
$S_m(0)$, where $S_m(\bold{q})=                       
\frac{1}{(Nx)^2} \sum_{ij} \bold{m}_{i}.\bold{m}_{j}e^{i\bold{q}.                       
  (r_{i}-r_{j})}$ (\textbf{q} are the wave vectors) along with quantum 
ferromagnetic structure factor $s(0)$ in Fig.~\ref{fig03}(b) and it shows the 
same transition. Due to strong coupling between itinerant carriers and auxiliary
fields the carriers are always align anti-parallel to the fields. For this reason
classical ferromagnetic structure factor $S(0)$ and quantum
ferromagnetic structure $s(0)$ behaves very similar to each other.

In order to figure out the correspondence between the ferromagnetism and the
metallicity we plot the temperature dependence of the resistivity for $n=0.2$ in
Fig.~\ref{fig04}(a). We calculate the {\it dc} limit of the optical conductivity by
using the Kubo-Greenwood formula~\cite{mahan,sanjeev1}.
At low temperature the system shows metallic behavior. The insulator-metal transition
coincides with the onset of ferromagnetism (see Fig.~\ref{fig04}(a)).
To check for finite size effect which remains a concern in the small system size
based Monte Carlo calculations, we show $s(0)$ with temperature for 
two system sizes L = 10 and 12 in Fig. \ref{fig04}(b). Our results indicate that
the curves are pretty similar to each other.

\begin{figure}[t]
\centering

\includegraphics[scale=0.35]{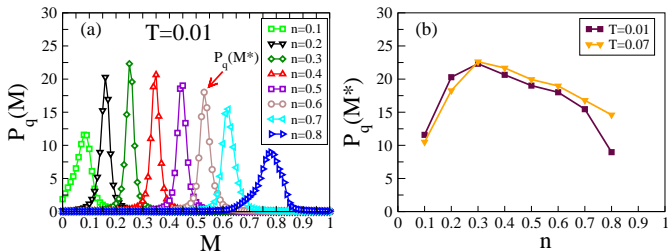}
\caption{Distribution of carrier moments for different carrier densities
(for $U=12$ and $x=0.25$): (a) The distribution of carrier moments $P_q(M)$ for
$U = 12$ case (using only $U = 12$ sites). $P_q(M)$ gets narrower up to $n=0.3$ 
and boardens thereafter. This shows that moments at $n=0.3$ are more uniform than 
other densities. (b) The peak of the distribution (defined as $P_q(M^*)$) 
for different $n$ shows that it is optimum for $n=0.3$.}
\label{fig05}
\end{figure}

\begin{figure}[t]
\centering

\includegraphics[scale=0.3]{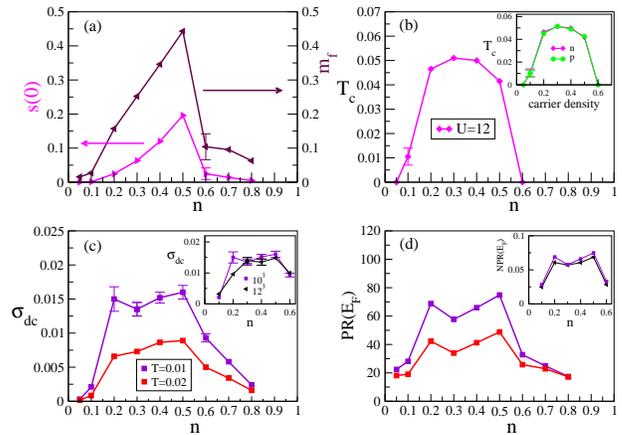}
\caption{Magnetic and transport properties ($U=12$ and $x=0.25$):
(a) Quantum ferromagnetic structure $s(0)$ obtained for $T=0.01$ plotted against 
electron densities. Uniform magnetization $m_{f}$ vs electron density at $T=0.01$ 
show one to one correspondence with $s(0)$ as it is proportional to the
square root of $s(0)$. (b) Ferromagnetic window with respect to the electron
density ($n$). The ferromagnetic transition temperature ($T_C$) shows 
optimization behavior. The inset shows the FM window for both electron
and hole densities. (c) {\it dc} conductivity calculated at $T=0.01$ show
metallicity at middle of the ferromagnetic window and depicts an 
insulator-metal-insulator (IMI) transition with respect to $n$.
(d) Participation ratio around Fermi energy PR($E_{F}$) shows that 
the states at the middle of ferromagnetic windows are more extended and
agrees well with conductivity results. Corresponding quantities for
two system sizes are compared in the inset of (c) and (d) }

\label{fig06}
\end{figure}

The moment distributions $P_q(M)$ for the different densities using only $U=12$
sites are shown in Fig.~\ref{fig05}(a) at $T = 0.01$. The moment distributions
get steeper and the peak value (defined as $P_q(M^*)$) increases up to $n=0.3$
and decreases thereafter. This shows that the moment distribution curve gets
broadened with a reduction in $P_q(M^*)$ value on both sides of $n=0.3$.
We plot the peak value of moment distribution curves $P_q(M^*)$ vs $n$
for $T = 0.01$ and $0.07$ in Fig.~\ref{fig05}(b). Next we will show that the
optimization of $P_q(M^*)$ is very similar to the optimized ferromagnetic,
conductivity and participation-ratio windows [see Fig.~\ref{fig06}].
This emphasize the fact that the optimum ferromagnetic $T_C$ is obtained 
for which the moment fluctuation at $U=12$ sites is minimal.

\begin{figure}[t]
\centering

\includegraphics[scale=0.3]{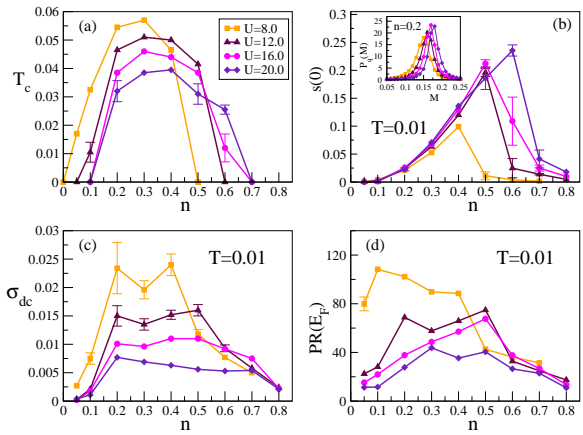}
\caption{Ferromagnetic windows for different $U$: (a) Ferromagnetic window 
shifts to right with $U$. (b) Ferromagnetic structure factor $s(0)$
vs. $n$ at $T=0.01$. $s(0)$ for all $U$ values show non-monotonic
behavior that agrees with the FM window.
(c) {\it dc} conductivity with respect to $n$ show IMI pattern for
different $U$ values.
(d)Participation ratio PR($E_F$) which is a measure of
delocalization also signifies that the states corresponding to FM order are
extended, though for higher $U$ values PR($E_F$) decreases. 
Inset in (b) shows that peak of local moment distribution shift towards the carrier
density value (for $n=0.2$) for higher $U$ and indicate the electrons are more
and more localized for larger $U$.}
\label{fig07}
\end{figure}

Now we analyze the quantum ferromagnetic structure factor for different carrier
densities $n$ at $T = 0.01$ using interacting $U = 12$ sites as mentioned in
Fig.~\ref{fig03}(a) and Fig.~\ref{fig05}(a). For $T = 0.01$ the quantum ferromagnetic
structure factor as shown in Fig.~\ref{fig06}(a) increases with carrier
density as expected due to the enhancement of the moments at the interacting sites
(see Fig.~\ref{fig05}(a)) and decreases sharply beyond $n  =0.5$. We also calculated
the system averaged uniform magnetization $m_f = \langle (n_{\uparrow} - n_{\downarrow}) \rangle$
for $U = 12$ sites and show that $m_f$ goes as square root of $s(0)$
[see Fig.~\ref{fig06}(a)]. The non-monotonic behavior of $s(0)$ (and $m_f$)
around $n = 0.5$ indicate that the ferromagnetic order vanishes for $n = 0.6$.
The ferromagnetic $T_C$ calculated for different $n$, shown in Fig~\ref{fig06}(b)
corroborate this fact. The ferromagnetic window exhibits optimum ordering at the
middle of the window. This emphasize the fact that, within our s-MC calculations,  
a minimum amount of carrier is essential to gain considerable kinetic energy 
to spin polarize the system. On the other hand, for higher carrier densities,
the magnetism is suppressed due to decrease in carrier mobility as the availability
of spatial interacting lattice sites decreases, which constrains the carrier
movement. In the inset of Fig.~\ref{fig06}(b) we show that the ferromagnetic window
and $T_C$ remains same for both hole and electron density calculations. The
spin-stiffness in spin wave calculations in diluted Hubbard model also
pose similar non-monotonic picture with carrier density~\cite{singh1}.
Similar results for magnetic impurities were also obtained in other spin
wave and MC calculations~\cite{alvarez,singh,pradhan1,bui,chakraborty}.

The non-monotonic ferromagnetic window signifies that the kinetic energy is
minimum at the edge of the ferromagnetic window. As a result, one expects a metallic
system at the center of the band and an insulating state at the edge for low
temperatures. In fact conductivity calculations for different $n$ plotted at $T=0.01$
(see Fig.~\ref{fig06}(c)) shows the same and depicts an insulator-metal-insulator
(IMI) transition with carrier density. This also establish the fact that the
mobility is minimum near the edge of the ferromagnetic window. The participation
ratio (PR) which is a measure of the localization is also calculated to
corroborate this fact. The participation ratio PR [$= 1/\sum_{i}(\psi_{l}^{i})^4$
where $\psi_{l}^{i}$ is the normalized quasiparticle wave function for $i$-th
site with $l$-th eigenvalue] provides a measure to see if the state is
localized or extended. PR($E_F$) which is the PR value around the
Fermi energy, is shown in Fig.~\ref{fig06}(d). Higher PR($E_{F}$) values for 
($n=0.2-0.5$) shows that the states in the middle of the FM window are
more extended, that agrees well with the ferromagnetic $T_C$
and conductivity data. In inset of Fig.~\ref{fig06}(c) and (d) we show that the
conductivity and normalized PR remains more or less same for two system sizes
(L = 10 and 12). 

\begin{figure}[t]
\centering

\includegraphics[scale=0.3]{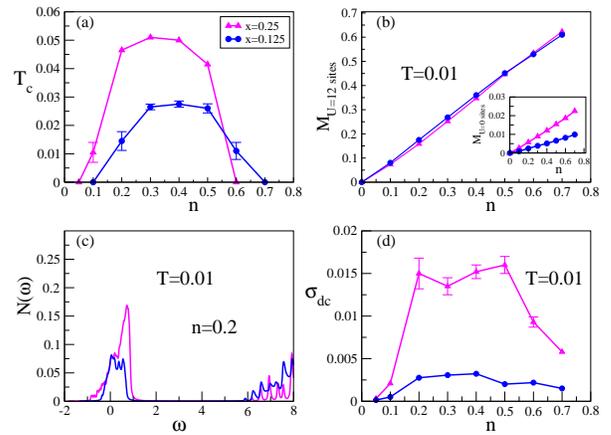}	
\caption{Comparison of magnetic and transport properties between $x=0.25$ and
$x=0.125$ (using $U = 12$): (a) FM window shift slightly and optimum $T_c$ decreases
considerably for $x=0.125$. (b) Average carrier moments at $U = 12$ sites for
both concentrations at different densities. Inset shows the average moments at
$U = 0$ sites.
(c) Density of states shows that impurity band gets narrower for smaller 
impurity concentration. This signify that the carriers are more 
localized for $x=0.125$ as compared to $x=0.25$.
(d) {\it dc} conductivity vs $n$ shows that the conductivity 
decreases for $x=0.125$.}

\label{fig08}
\end{figure}

\section{Comparison of ferromagnetic windows for different $U$}

Next, we explore the ferromagnetic window for four different $U$ values. The 
ferromagnetic window in Fig.~\ref{fig07}(a) shifts to the right with increasing
$U$. This is because the carriers are more localized for larger U 
values [see in inset of Fig. \ref{fig07}(b)] and enhances the carriers mobility 
among the interacting sites due to availability of more interacting lattice sites
beyond $n=0.5$. This analysis is true for higher
density edge of the impurity band. In fact, in large $U$ ($>> BW$) limit ferromagnetic
window is expected to span up to $n = 1.0$ like undiluted Kondo lattice
model~\cite{yunoki,pradhan}. At the same time ferromagnetism in
lower part of the FM window gets depleted for stronger localization of carriers
that remain localized far apart on impurity sites. The quantum ferromagnetic structure
factor for different $U$ values (see Fig. \ref{fig07}(b)) at low temperature ($T=0.01$)
also shows the non-monotonic behavior similar to $U=12$ case. Here,
for all $U$, the maximum $s(0)$ is found for the particular carrier density beyond
which the ground state is paramagnetic.
In Fig.~\ref{fig07}(c) the conductivity results show that the insulator-metal-insulator
(IMI) pattern also remains intact for all $U$. With increasing $U$ the conductivity
decreases, mainly at the middle of FM window, as carriers are more localized at
larger $U$ values. In addition PR($E_{F}$) in
Fig.~\ref{fig07}(d) supports the localization-delocalization-localization
pattern with the electron density which is also obtained from the ferromagnetic and
conductivity calculations.

\section{Comparison of Ferromagnetic Windows between $x =0.25$ and $x =0.125$}

In addition to $x=0.25$, we performed systematic calculations to unveil the
ferromagnetic window for $x=0.125$ case. In Fig.~\ref{fig08}(a) we compare
the ferromagnetic windows for $x=0.25$ and $0.125$ using $U=12$. Here the FM
window with respect to $n$ shifts to the right, but $T_C$ reduces considerably.
The average local moments on $U=12$ sites remain same for both $x$
[see Fig.~\ref{fig08}(b)]. This indicate that the induced moment at $U=0$
sites would decrease for smaller $x$. In fact as we decrease
the $x$ the resulted moments at $U=0$ sites decreases [see the inset of
Fig.~\ref{fig08}(b)]. Although moments for $U=12$ sites are the same, the density
of states in Fig.~\ref{fig08}(c) shows that the carriers are more
localized as the impurity band gets narrower for $x=0.125$. This is
due the increase of the distance between interacting sites in $x=0.125$
as compared to $x=0.25$. This is one of the reason for the drop in $T_C$ value.
In Fig. \ref{fig08}(d) the reduction of conductivity with $x$ agrees with
the fact that the mobility of carriers decline for $x=0.125$.

\section{Conclusions}

In summary, based on spin-fermion model, we show that the ferromagnetism is
favored for low density of carriers, which is concomitant to the experimental
results. In a non-perturbative limit ($U \sim BW$) our analysis shows that
the density of the itinerant carriers, confined to  the impurity band,
that decides the kinetic energy of the system, plays an important role in
determining the carrier spin polarization. The ferromagnetic
ordering temperature shows an optimization behavior with the carrier density.
We have provided a systematic study of carrier spin dependent transport properties
of the carriers over the whole carrier density range. A insulator-metal-insulator
transition is observed across the ferromagnetic window. Due to strong coupling
between itinerant carriers and auxiliary fields the polarization of auxiliary 
fields follows the property of carrier spin polarization very closely. Thus
our results are significant for the understanding of ferromagnetism in diluted
magnetic semiconductors.

\noindent \\
Acknowledgment: We acknowledge use of Meghnad2019 computer cluster at SINP and our
discussion with S. K. Das and A. Mukherjee.

\end{document}